\documentclass[11pt]{article}
\usepackage{graphicx}                  
\usepackage{amssymb, tensor}
\usepackage{amsmath, mathptmx}
\numberwithin{equation}{section}
\usepackage{multicol}
\usepackage{array}
\newcolumntype{C}[1]{>{\centering\arraybackslash}m{#1}}
\usepackage{epstopdf}
\usepackage{caption}
\usepackage{subcaption}
\usepackage{mathrsfs}
\usepackage{grffile}
\usepackage{floatrow}
\usepackage[amssymb]{SIunits}
\usepackage{floatflt} 
\usepackage[toc,page]{appendix}
\usepackage[listings,theorems]{tcolorbox}
\usepackage[section]{placeins}
\usepackage{wrapfig}
\usepackage{geometry}
\usepackage[utf8]{inputenc} 
\usepackage[amssymb]{SIunits}
\usepackage{geometry}
\usepackage{setspace}
\usepackage{physics}
\usepackage{titlecaps}
\usepackage{titlesec}
\usepackage[bookmarks]{hyperref}
\usepackage{titlesec}
\usepackage{authblk}
\usepackage{tikz}
\titleformat*{\section}{\large\bfseries}
\titleformat*{\subsection}{\normalfont\bfseries}
\titleformat*{\subsubsection}{\small\bfseries}

\title{\textbf{Gravitational redshift revisited: inertia, geometry, and charge}}
\author{}
\author{Johannes Fankhauser\thanks{Institute for Theoretical Physics, University of Innsbruck, Austria. \href{mailto:johannes.j.fankhauser@gmail.com}{ johannes.j.fankhauser@gmail.com}}~ and James Read\thanks{Faculty of Philosophy, University of Oxford, UK. \href{mailto:james.read@philosophy.ox.ac.uk}{ james.read@philosophy.ox.ac.uk}}}

\date{}                                           

\begin{document}
\maketitle

\renewcommand{\abstractname}{\vspace{-3\baselineskip}}
\begin{abstract} 

Gravitational redshift effects undoubtedly exist; moreover, the experimental setups which confirm the existence of these effects---the most famous of which being the Pound-Rebka experiment---are extremely well-known. Nonetheless---and perhaps surprisingly---there remains a great deal of confusion in the literature regarding what these experiments really establish. Our goal in the present article is to clarify these issues, in three concrete ways. First, although (i) \cite{Brown2016} are correct to point out that, given their sensitivity, the outcomes of experimental setups such as the original Pound-Rebka configuration can be accounted for using solely the machinery of accelerating frames in special relativity (barring some subtleties due to the Rindler spacetime necessary to model the effects rigorously), nevertheless (ii) an explanation of the results of more sensitive gravitational redshift outcomes \emph{does} in fact require more. Second, although typically this `more' is understood as the invocation of spacetime curvature within the framework of general relativity, in light of the so-called `geometric trinity' of gravitational theories, in fact curvature is not \emph{necessary} to explain even these results. Thus (a) one can often explain the results of these experiments using only the resources of special relativity, and (b) even when one cannot, one need not invoke spacetime curvature. And third: while one might think that the absence of gravitational redshift effects would imply that spacetime is flat (indeed, Minkowskian), this can be called into question given the possibility of the cancelling of gravitational redshift effects by charge in the context of the Reissner-Nordström metric. This argument is shown to be valid and both attractive forces as well as redshift effects can be effectively shielded (and even be repulsive or blueshifted, respectively) in the charged setting. Thus, it is not the case that the absence of gravitational effects implies a Minkowskian spacetime setting.


\end{abstract}

\setlength\columnsep{7mm}
\begin{multicols}{2}

\tableofcontents

\section{Introduction}

In 1911, Einstein foresaw a phenomenon thereafter known as `gravitational redshift' \cite{einstein1911einfluss}. His thought experiment initiated the revolutionary idea that mass `warps' space and time. There does, however, remain---even after over a century of study---some confusion in the literature regarding what can be inferred legitimately about the nature of space and time on the basis of the results of gravitational redshift experiments.
Our goal in this article is to clarify this issue, in three ways. First, although (i) \cite{Brown2016} are correct to point out that, given their limited sensitivity, the outcomes of experimental setups such as the original configuration of \cite{pound1960apparent} can be accounted for using solely the machinery of accelerating frames in special relativity (barring some subtleties due to the Rindler spacetime necessary to model the effects rigorously), nevertheless (ii) an explanation of the results of more sensitive gravitational redshift outcomes \emph{does} in fact require more. Second, although typically this `more' is understood as the invocation of spacetime curvature within the framework of general relativity, in light of the so-called `geometric trinity' of gravitational theories, in fact curvature is not \emph{necessary} to explain even these results. Thus (a) one can often explain the results of these experiments using only the resources of special relativity, and (b) even when one cannot, one need not invoke spacetime curvature. And third: while one might think that the absence of gravitational redshift effects implies that spacetime is flat, this can be called into question given the possibility of the cancelling of gravitational redshift effects by charge in the context of the Reissner-Nordström metric. This argument is shown to be valid and both attractive forces as well as redshift effects can be effectively shielded (and even be repulsive or blueshifted, respectively) in the charged setting. Thus, it is not the case that the absence of gravitational effects implies a Minkowksian spacetime setting.


The structure of the article is this. In \S\S\ref{sec:redshift}--\ref{sec:equiv}, we derive and discuss the gravitational redshift effect in three ways: (i) from the framework of general relativity (GR), (ii) using the equivalence principle, and (iii) from energy conservation principles. We then compare the results, and find them to be different; this allows us to be explicit about when one can account for the outcomes of gravitational redshift experiments using only the resources of special relativity, and when one cannot, thereby making good on our first self-declared goal as presented above. In \S\ref{sec:torsion}, we introduce the geometric trinity of gravitational theories---which trade the spacetime curvature of GR for either torsion (in the case of the theory known as `teleparallel gravity') or spacetime non-metricity (in the case case of the theory known as `symmetric teleparallel gravity')---and show that by invoking this trinity of theories, one need \emph{not} appeal to spacetime curvature in order to explain even the exact results of gravitational redshift experiments beyond first order. Along the way, we demonstrate the falsity of some recent claims in the literature that gravitational redshift experiments provide \emph{direct} evidence for spacetime torsion; together, all this allows us to make good on our second self-declared goal as presented above. In \S\ref{sec:charge}, we examine effects on the redshift due to charge with some remarks on the relationship between GR and electromagnetism, and the possibility of cancelling locally gravity with charge; ultimately we find that one can shield both effective gravitational forces and redshift effects in the Reissner-Nordström metric; so, the absence of gravitational redshift effects does not imply that spacetime is Minkowskian; this makes good on our third self-declared goal as presented above. We close in \S\ref{sec:conclusion}.

\section{Gravitational redshift}\label{sec:redshift}

It is a straightforward exercise to derive the relative shift in proper time measured by two clocks in a given gravitational field with metric $g_{ab}$. Since we will employ in the following sections some alternative approximate approaches to deriving the gravitational redshift result, we first present the exact and most general derivation from general relativity, variants of which are standard fare (see for example, \cite[p.~136]{wald2010general}). 

\begin{figure}[H]
\centering
\includegraphics[width=0.9\linewidth]{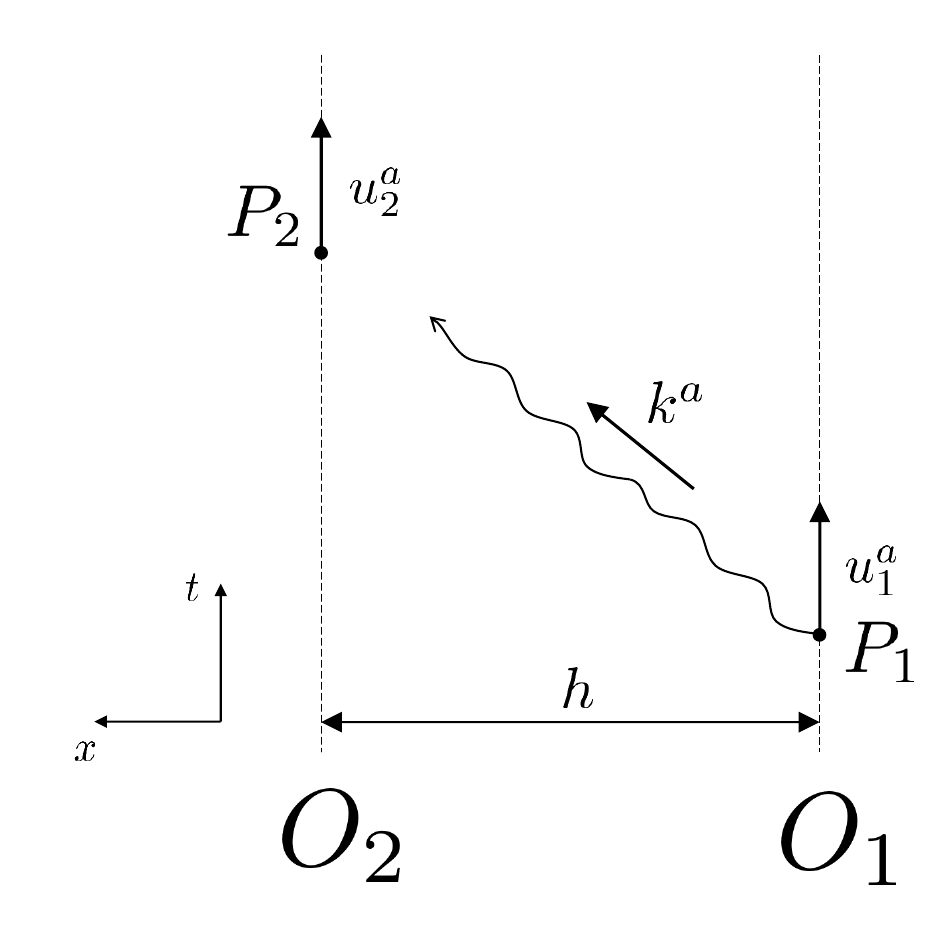}
\caption[]{Two observers at different heights experience a time dilation effect in Earth's gravitational field. Emitter $O_1$ on the surface of the Earth sends a train of electromagnetic pulses from point $P_1$ with energy momentum 4-vector $k^a$ to a receiver $O_2$, placed at point $P_2$, at height $h$ above $P_1$. We assume $O_1$ and $O_2$ are stationary, i.e. their 4-velocities $u_1^a$ and $u_2^a$ are tangential to the  Killing field $\xi^a=\left(\frac{\partial}{\partial t}\right)^a$.}
\label{fig:redshift}
\end{figure}

An emitter $O_1$ on the surface of the Earth sends a train of electromagnetic pulses from point $P_1$ with energy momentum 4-vector $k^a$ to a receiver $O_2$, placed at point $P_2$, at height $h$ above $P_1$. We assume the two observers $O_1$ and $O_2$ to be stationary, which is to say that their 4-velocities $u_1^a$ and $u_2^a$ are tangential to the stationary Killing field $\xi^a=\left(\frac{\partial}{\partial t}\right)^a$. Since the 4-velocities of the two observers are unit vectors pointing in the direction of $\xi^a$, we have $u_1^a=\left.\frac{\xi^a}{\sqrt{-\xi^b\xi_b}}\right\rvert_{P_1}$ and $u_2^a=\left.\frac{\xi^a}{\sqrt{-\xi^b\xi_b}}\right\rvert_{P_2}$. The lengths $\sqrt{-\xi^b\xi_b}=\sqrt{-g_{bc}\xi^b\xi^c}$ are obtained by contraction with the metric. We let the observers $O_1$ and $O_2$, whose clock rates we wish to compare, describe their world-lines. The difference in the world-lines' lengths in spacetime consequently determines the amount of gravitational redshift. Figure \ref{fig:redshift} illustrates the thought experiment.

Recall that for a given energy-momentum 4-vector $p^a=mu^a$ of a particle, with respect to a local inertial frame, the energy observed by an observer that moves with 4-velocity $v^a$ is 
\begin{equation}
\label{energy}
E=-p^av_a.\footnote{In particular, if $u^a=v^a$, i.e. the particle's 4-velocity aligns  with the observer's, then $E=-mv^av_a=mc^2$ (choosing the metric signature to be $(-,+,+,+)$).}
\end{equation}
Therefore, for the frequency $\nu_i$ of the photon observed by $O_i$, which moves with 4-velocity $u_i^a$, we find the relation $h\nu_i=E_k=\left.-k_au^a_i\right\rvert_{P_i}$ (cf.~\eqref{energy}), where $E_k$ is the energy of the photon.\footnote{Here $h$ is just a constant that relates the energy of a photon to its frequency, and there is nothing quantum in this argument.} By definition of the vector field $\xi^a$, we have $\left.\xi_a\xi^a\right\rvert_{P_i}=\left.g_{00}\right\rvert_{P_i}$ since $\xi^a$ has vanishing spatial components. It would involve a fair amount of work to derive the gravitational redshift by finding the geodesic equation. However, this can be avoided by taking advantage of a useful proposition. Light travels on null geodesics (in the geometrical optics approximation, i.e.~the spacetime scale of variation of the electromagnetic field is much smaller than that of the curvature: see e.g.~\cite[p.~571]{mtw}), from which it follows that the inner product $k_a\xi^a$ is constant along geodesics, that is $\left.k_a\xi^a\right\rvert_{P_1}=\left.k_a\xi^a\right\rvert_{P_2}$.\footnote{For a detailed proof see for instance, \cite[p.~442]{wald2010general}}

Spacetime around Earth (if considered as generated by a point mass $M$ at $r=0$) can be modelled by the Schwarzschild metric
\begin{align}
\label{Schwarzschild}
ds^2=g_{\mu\nu}dx^{\mu}dx^{\nu}=&-\left(1-\frac{r_S}{r}\right)c^2dt^2 \nonumber\\
&+\left(1-\frac{r_S}{r}\right)^{-1}dr^2\nonumber \\
&+r^2(d\vartheta^2+\sin^2\vartheta d\varphi^2),
\end{align} where 
\begin{equation}
r_S= \frac{2GM}{c^2}
\end{equation} is the so-called Schwarzschild radius, $r$ the distance from the Earth's centre, $G$ the gravitational constant, $c$ the speed of light, and $M$ the mass of the Earth. This yields
\begin{align}
\label{redshift}
\frac{\nu_1}{\nu_2}&=\frac{\left.\sqrt{-\xi^b\xi_b}\right\rvert_{P_2}}{\left.\sqrt{-\xi^b\xi_b}\right\rvert_{P_1}}=\frac{\sqrt{1-\frac{2GM}{c^2r_2}}}{\sqrt{1-\frac{2GM}{c^2r_1}}} \nonumber\\
&\approx 1 + \frac{GM}{c^2}\left(\frac{1}{r_1}-\frac{1}{r_2}\right) 
\approx 1+ \frac{gh}{c^2}, 
\end{align} or
\begin{equation}
\frac{\Delta\nu}{\nu}\approx \frac{GM}{c^2}\left(\frac{1}{r_1}-\frac{1}{r_2}\right),
\end{equation} with $g:=\frac{GM}{r_1^2}$ the gravitational constant at $r_1$, $\nu=\nu_1$, $\Delta \nu=\nu_1-\nu_2$, and $r_2-r_1=h$. For the last approximation in the second last line we have used $\frac{1}{r_1}-\frac{1}{r_2} = \frac{r_2-r_1}{r_2r_1}\approx \frac{h}{r_1^2}$ if $r_1\approx r_2$ and $r_1,r_2 \gg h$.
 
Experimental tests of the gravitational redshift were first conducted by Cranshaw, Schiffer and Whitehead in the UK in 1960 \cite{cranshaw1960measurement}. It was not clear whether significant conclusions could be drawn from their results. In the same year, the experiments by Pound and Rebka in Harvard successfully verified the gravitational redshift effect \cite{pound1960apparent}.

\section{Uniformly accelerated frames and the equivalence principle}\label{sec:EP}

Einstein's equivalence principle (also called the weak equivalence principle) assumes that any experiment in a uniform gravitational field yields the same results as the analogous experiment performed in a frame removed from any source of gravitational field but moving in uniform accelerated motion with respect to an inertial frame 
\cite{norton1985einstein}.\footnote{Note that \cite{Brown2016} use `Einstein equivalence principle' to refer to what is often called the `strong equivalence principle'. For further recent discussion of equivalence principles, see \cite{Lehmkuhl}.} 

However, it is clear that Einstein was well aware of the mere linearly approximate validity of the equivalence principle when he wrote that
\begin{quote}
we arrive at a principle [the equivalence principle] which, if it is really true, has great heuristic importance. For by theoretical consideration of processes which take place
relative to a system of reference with uniform acceleration, we obtain information as to the behaviour of processes in a homogeneous gravitational field. [...] It will be shown in a subsequent paper that the gravitational field considered here
is homogeneous only to a first approximation. \cite[p.~900]{einstein1911einfluss}
\end{quote} 
The principle, thus, holds only in a `small neighbourhood' of a point-like observer. 
Nonetheless, a treatment of the redshift effect in a uniform static gravitational field proves instructive, insofar as it shows that certain consequences of GR can be explained without resorting to effects such as spacetime curvature (this, indeed, is the central lesson of \cite{Brown2016}). Dealing with uniform accelerations in order to derive the gravitational redshift, however, is a delicate business, and we shall see that the field, resulting from uniform (proper) acceleration, is not \textit{uniform} if we demand a constant (proper) distance between emitter and observer! This, of course, is a familar lesson regarding Rindler frames (i.e.,~uniformly accelerating frames) in special relativity): see e.g.~\cite[ch.~9]{Read} for further discussion.

\begin{figure}[H]
\centering
\includegraphics[width=0.45\linewidth]{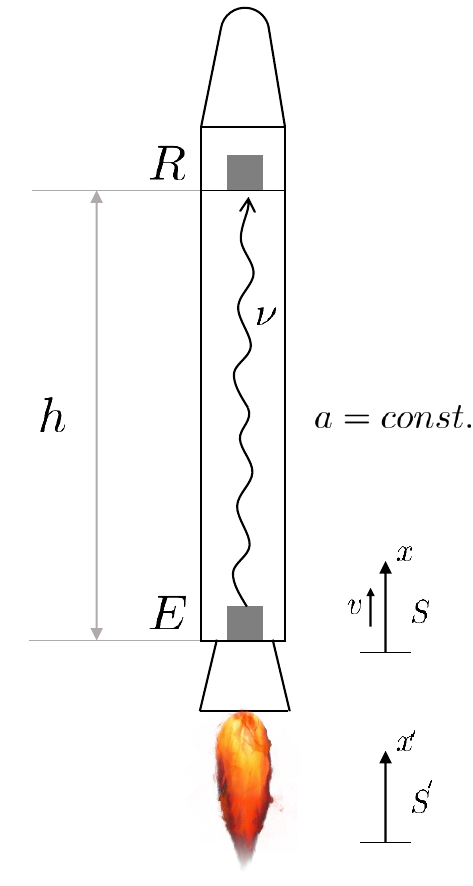}
\caption{The gravitational redshift experiment in a uniformly accelerated spaceship. The redshift effect can be explained by the equivalence principle---\emph{to first order}.}
\label{fig:Pound-Rebka equivalence}
\end{figure}

We consider a spaceship that is uniformly accelerated. An emitter $E$ and receiver $R$ inside the spaceship, separated by a height $h$, compare frequencies of  signals ascending the spaceship. For an illustration, see Figure \ref{fig:Pound-Rebka equivalence}.
As in the derivation of the gravitational redshift from the Schwarzschild metric, we let the observers describe their world-lines. It suffices to consider only one spatial dimension $x$. Acceleration $a$ is measured in an inertial frame $S$ with  momentary velocity $v$ relative to the inertial frame $S'$ outside the spaceship, inside of which the acceleration is measured to be $a'$.\footnote{It is implicitly assumed that the proper time of co-moving clocks depends only on velocity and is independent of acceleration. This assumption is often called the `clock hypothesis' (see for example, \cite[Section~3]{Brown2016}).} Relativistic transformation of 3-acceleration gives
\begin{equation}
\label{acc}
a=\gamma^3a',
\end{equation} where $\gamma=\frac{1}{\sqrt{1-\frac{v^2}{c^2}}}$ is the Lorentz factor.\footnote{To find the transformation of acceleration, one has to differentiate the spatial coordinates of the Lorentz transformation with respect to the time coordinates to first find the 3-velocity transformation (velocity-addition formula). Another differentiation of the velocities yields the transformation law for 3-acceleration.} 

Note that the acceleration of the spaceship needs to be measured in the (momentary) inertial frame with instantaneous velocity $v$ such that $a'=\frac{dv}{dt}$ (proper acceleration). With respect to the accelerated frame, sure enough, the ship's acceleration is zero. However, the principle of relativity---the requirement according to which the laws of physics take the same form in any inertial frame---no longer holds in accelerated, hence non-inertial, frames. Therefore, as expected, the two observers in the spaceship are going to feel a (pseudo)force $F=m_0a$, where $m_0$ is the rest mass (invariant mass) of an object in the spaceship.  

We want the (proper) acceleration $a$ of the spaceship to be constant. The right hand side of \eqref{acc} is equal to $\frac{d}{dt} \left(\gamma v\right)$. Since $a$ is constant we integrate \eqref{acc} twice to find the trajectory---a so-called Rindler hyperboloid---of a uniformly accelerated point body as observed in the inertial frame $S'$:

\begin{equation}
x(t)=\frac{c^2}{a}\sqrt{1+\left(\frac{at}{c}\right)^2}+C,
\end{equation} with $C$ a constant from integration. The second constant from the first integration was set to zero such that $v(0)=0$. Without loss of generality we can also set $C=0$. The result represents a hyperbolic path in Minkowski space, i.e.

\begin{equation}
x^2-c^2t^2= \frac{c^4}{a^2},
\end{equation} from which the term `hyperbolic motion' is derived. We assume the back of the spaceship be subject to this motion. Note that $\dot{x}\overset{t\rightarrow \infty}{\rightarrow}c$, as expected. 

We recover uniform acceleration in the Newtonian sense for $t\ll1$. That is,
\begin{align}
x(t)&=x_0+\frac{at^2}{2},
\end{align} with $x_0=c^2/a$ the position at $t=0$. 

\begin{figure}[H]
\centering
\includegraphics[width=0.8\linewidth]{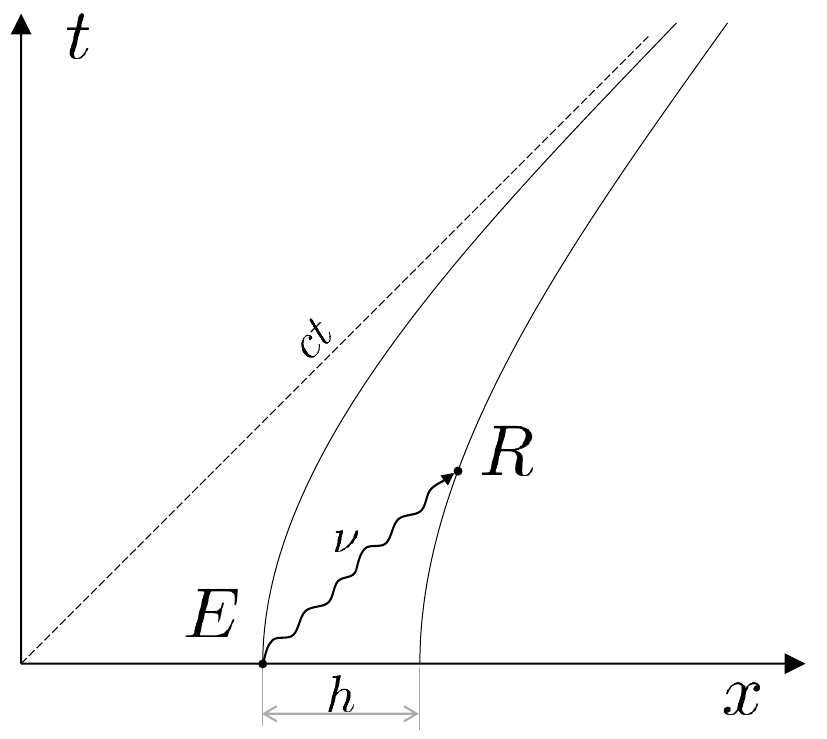}
\caption{The world-lines of emitter $E$ and receiver $R$ are Rindler hyperbolae when experiencing constant proper acceleration.}
\label{fig:Rindler hyperbolae}
\end{figure}

For an exact derivation, it would lead to inconsistencies to assume that the emitter and receiver traverse the same Rindler hyperboloid with only an additional spatial distance $h$ in the coordinate $x$ between them. For if we maintain a constant height between $E$ and $R$ relative to the inertial observing frame $S'$, then length contraction, as predicted in special relativity, will stretch the spaceship and eventually tear it apart (cf.~the spaceship paradox in \cite{dewan1959note} and \cite[ch.~9]{bell1987speakable}).  This is key. As was also pointed out by \cite{Alberici2006}, assuming the gravitational acceleration to be the same for the top and bottom observers leads to all kinds of paradoxes. Most notably, it is not possible in this case to define a globally freely falling inertial frame because the corresponding metric would lead to a non-vanishing Riemann tensor, and hence curvature! (Cf.~our discussion of Synge's argument in \S\ref{sec:torsion}.) The receiver $R$ in the bow lying higher by  height $h$ with respect to the emitter $E$ must follow the hyperboloid 
\begin{equation}
x^2-c^2t^2= \left(\frac{c^2}{a}+h\right)^2,
\end{equation} for the proper height (relative to $S$) to be constant.  These are the two desiderata to simulate reasonably the gravitational redshift by uniform acceleration: first, the ship must have a constant acceleration; and second, the ship must have a constant proper height. The world-lines of emitter and receiver are denoted in Figure \ref{fig:Rindler hyperbolae}. 
Due to relativistic length contraction, the receiver's proper acceleration needs to be slightly greater. By comparing the two hyperbolae it immediately follows that the acceleration $g_R$ of the receiver is related to the emitter's acceleration $g_E$ by
\begin{equation}
g_R=\frac{g_E}{1+\frac{g_Eh}{c^2}}.
\end{equation} (Compare also the treatment and related paradoxes in \cite{fabri1994paradoxes}.) Therefore, the gravitational field is not constant over the extended region of the spaceship. That is, however, not a surprise, for we would not expect the equivalence principle to hold globally in the first place.   
Further, it follows that proper time intervals along two different Rindler hyperbolae between two events having the same coordinate velocity are in a fixed proportion, 
\begin{equation}
\frac{\tau_R}{\tau_E}=\frac{g_E}{g_R}=1+\frac{g_Eh}{c^2},
\end{equation} yielding the exact gravitational redshift formula for uniform acceleration. Alternatively, we can write
\begin{equation}
\nu_R=\frac{\nu_E}{1+\frac{g_Eh}{c^2}}=\nu_E\left(1-\frac{g_Rh}{c^2}\right)
\end{equation} for the corresponding observed frequencies, to highlight the dependence on the two different proper accelerations of emitter and receiver (cf.~also the results in \cite{Alberici2006}). From the preceding derivations we readily find for the (Rindler) metric of an accelerated frame
\begin{equation}
ds^2=g_{\mu\nu} dx^{\mu}dx^{\nu} = \left(1+\frac{g_Ex}{c^2}\right)^2c^2dt^2-dx^2.
\end{equation} Thus, the gravitational redshift according to this metric reads
\begin{equation}
\label{egn:Rindler shift}
\frac{\nu_E}{\nu_R}=\frac{\Delta t_{x=h}}{\Delta t_{x=0}}=\frac{\sqrt{g_{00}}\rvert_{x=h}}{\sqrt{g_{00}}\rvert_{x=0}} =1+\frac{g_Eh}{c^2},
\end{equation}which is consistent with the first order approximation of the gravitational redshift from the Schwarzschild metric in \eqref{redshift}. Clocks at $E$ and $R$, whose rates one wishes to compare, are permitted to describe their world-lines, i.e.~Rindler hyperbolae, with respect to the inertial frame, and the value for the redshift is obtained by comparing the lengths of their world-lines in spacetime. Therefore, the treatment here is exact. The Rindler metric is, in fact, a solution to the vacuum Einstein field equations and has vanishing curvature (this should be obvious, since it is simply Minkowski spacetime in an accelerating frame). Note that since the redshift effect according to the Rindler metric depends on the absolute height $x$, it only coincides with the classical Doppler shift formula---which exactly equals  \eqref{egn:Rindler shift}---at the time when the space ship launches, i.e.~the emitter is at $x=0$, and the receiver at $x=h$. In this case the proper acceleration is equal to the gravitational acceleration on the surface of the earth, i.e. $g_E=g$. 

It is worth mentioning that the relativistic Doppler shift is yet another way to arrive at the gravitational redshift to first order. There, we have

\begin{equation}
\frac{\nu_E}{\nu_R}=\frac{\sqrt{1+\frac{v}{c}}}{\sqrt{1-\frac{v}{c}}}\approx 1+\frac{g_Eh}{c^2},
\end{equation} where $v=\frac{g_E h}{c}$ the velocity of the receiver at the time when the photon reaches it. 

\begin{figure}[H]
\centering
\includegraphics[width=\linewidth]{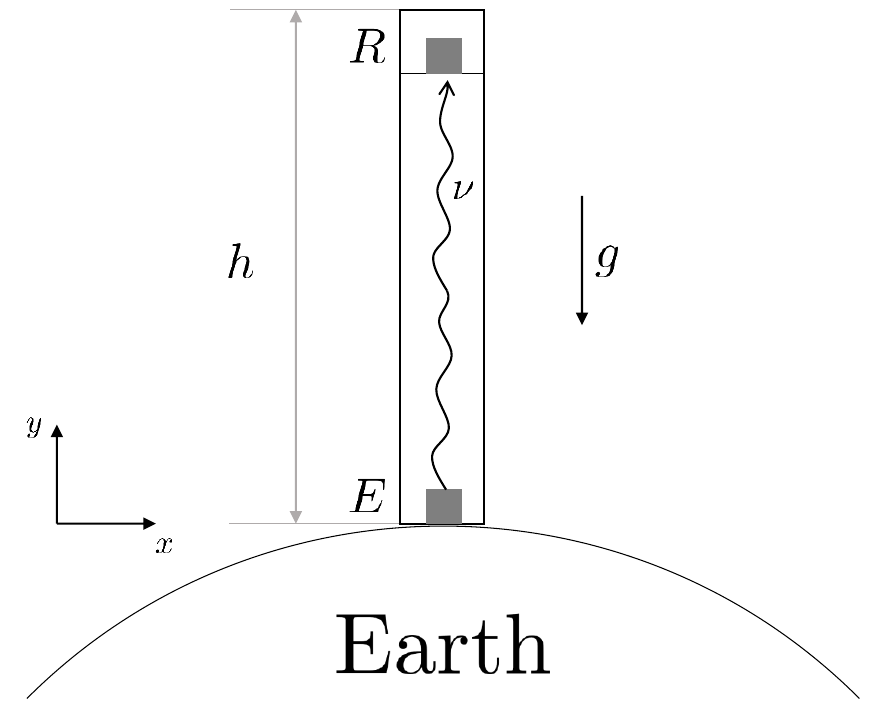}
\caption{The Pound-Rebka experiment. Receiver $R$ measures a lower frequency of the photon than what it was when emitted at $E$.}
\label{fig:Pound-Rebka}
\end{figure}

In experiments such as those of Pound and Rebka which were used to confirm gravitational redshift, the emitter sends a signal at equal intervals on a clock at the surface of the Earth. The receiver measures the time interval between receipt of the signals on an identical clock at height $h$ (see Figure \ref{fig:Pound-Rebka}).
Merely when the experiment is taken to be at rest in a Rindler frame, the equivalence principle implies that the relation between the clock times of emitter and receiver must be the same as if a spaceship were to accelerate vertically upwards in free space, as shown in Figure \ref{fig:Pound-Rebka equivalence}. The signals at the back are received at longer intervals than they are emitted because they are catching up with the accelerated bow of the spaceship and thus exhibit a Doppler shift. Note that the equivalence principle is local. Thus, in a field like that of the Earth it holds only approximately (to first order) for a small spacetime region.

\section{Equivalence and gravitational redshift}\label{sec:equiv}

Although GR is a well-established framework, it often occurs that its application amounts to an analysis that renders conclusions equivocal. This, in particular, happens to be the case for gravitational redshift. For instance, Brown and Read comment on the gravitational redshift effect as follows:

\begin{quote}
The second possible misconception [regarding general relativity] relates to the notion that gravitational redshift experiments provide evidence for spacetime curvature. They do, but contrary to what is claimed in some important modern textbooks on GR, a single gravitational redshift experiment does not require an explanation in terms of curvature. Rather, it is only multiple such experiments, performed at appropriately different locations in spacetime, that suggest curvature, via the notion that inertial frames are only defined locally [...] This ``redshift'' effect follows directly from the claim that the emitter and absorber are accelerating vertically at a rate of $g$ $m/s^2$ relative to the (freely falling) inertial frames. \cite[pp.~327, 329]{Brown2016}
\end{quote}

\noindent Here, Brown and Read assume the `redshift' effect to be independent of `tidal effects' (which is what they refer to as curvature). We have in fact already shown such a derivation is limited and does not fully account for gravitational redshift. Curvature effects are relevant in a single redshift experiment as outlined above in the most general derivation. Moreover, as we have seen, assuming both emitter and absorber to accelerate at the same rate is impossible given the two desiderata mentioned. 
However, they acknowledge there is nonetheless a connection between spacetime curvature and redshift experiments. This connection, for Brown and Read, amounts to the fact that redshift experiments carried out at different places on the surface of the Earth reveal `geodesic deviation' due to the spherical shape of the planet. That is, relative to a global freely falling frame at the site of one redshift experiment, a freely falling frame at another site is not moving inertially. Multiple gravitational redshift experiments thus require for their joint explanation the rejection of the global nature of inertial frames. Brown and Read maintain it is only geodesic deviation that reveals curvature. However, we have now seen that one (sufficiently sensitive!) experiment is in fact sufficient to detect tidal effects of Earth's gravitational field, and therefore curvature, after all (at least modulo the issues to be discussed in the following section). 

What Brown and Read deem to be a misconception, that is that
\begin{quote}
[a]n explanation for the results of a single gravitational redshift experiment of Pound–Rebka type will appeal to a notion of spacetime curvature \cite[p.~330]{Brown2016},
\end{quote}
is indeed one. However, this results not from an absence of curvature. Rather, since the Pound-Rebka experiment was solely designed to verify the first order effects predicted by GR, in this case a derivation via accelerated frames gives the desired result. 

Brown and Read's proposal holds if the gravitational field of the Earth is assumed to be uniform---that is, independent of the radial distance from the centre of the Earth, and also if $\frac{gh}{c^2}\ll1$. In experiments involving larger spatial separations or stronger gravitational field variations, it is necessary to use the exact Schwarzschild solution of GR. By means of fully formed GR, of course, all approximations are bound to disappear. Incidentally, the ratio between the exact gravitational redshift and the first order approximation  amounts to about 0.7$\%$---which is below the measurement accuracy of the Pound-Rebka experiment (typically around $1\%$  \cite{pound1960apparent}). However, more accurate experiments performed after that of Pound and Rebka are indeed able to measure gravitational redshift to a precision beyond the first order effect (see, for instance, the hydrogen maser clock tests with a height difference of about $10,000$ km by \cite{Vessot-accurate-shift}---the experiment tested gravitational redshift to 0.007$\%$ accuracy.) Thus, for the high precision measurements Read and Brown's account is insufficient to explain the effects of gravitational redshift in terrestrial experiments by appealing to the equivalence principle only. 


So: if the equivalence principle is to be used to explain the gravitational redshift, then it is important to realise that this can only be done to first order. In addition, the quantitative results of Pound-Rebka can indeed be justified without appealing to spacetime curvature, but one should be aware that a complete theoretical description has to take into account the inhomogeneous gravitational field of Earth. After all, more sophisticated experiments with higher accuracy than those used by Pound and Rebka do measure effects due to curvature in a single redshift experiment.\footnote{Note that for the Schwarzschild metric $R=0$ and $R_{\mu\nu}=0$, but not all entries of the Riemann curvature tensor $R\indices{^{\mu}_{\nu\rho\sigma}}$ vanish.} Although our considerations do not inhibit the successful comparison of the results of the Pound-Rebka experiment with first order calculation because higher order effects are beyond their measurement accuracy, they show that the qualitative explanation of the result does require one to invoke spacetime curvature and an exact treatment of accelerations in special relativity to model gravitational redshift with the equivalence principle. 

\section{Redshift and torsion}\label{sec:torsion}

Having established that first-order gravitational redshift effects do not require an explanation in terms of geometrical properties of spacetime such as curvature, let us turn now to the question of whether one needs spacetime curvature to explain the results of gravitational redshift experiments even \emph{beyond} first order.

\subsection{The geometric trinity}

What we have established up to this point is this: although \cite{Brown2016} are correct that one can account for the experimental results such as that of Pound and Rebka using only the resources of an accelerating frame in special relativity, the full explanation of the results of experiments of this kind beyond first order requires further resources, e.g.~recourse to spacetime curvature. Even granting this, however, it is important to recognise that although appeals to curvature might be \emph{sufficient} to explain such effects, they are not \emph{necessary}. The reason for this is that general relativity forms but one corner of a `geometric trinity' of gravitational theories, all of which are dynamically equivalent (in the sense that their Lagrangians are equivalent up to boundary terms\footnote{Whether this means that the theories are \emph{empirically} equivalent is a subtle business, and depends upon how the boundary terms by which the theories differ are treated---see \cite{wolf2023respecting} for discussion.}), but in each of which gravity is a manifestation of a different geometric property of spacetime: curvature in the case of general relativity, torsion in the case of `teleparallel gravity' (TPG), and non-metricity in the case of `symmetric teleparallel gravity' (STGR). For a review of the geometric trinity, see \cite{Jimenez}; in what follows we focus on the case of torsion and TPG.

We begin by recalling some details regarding spacetime torsion. The torsion tensor $T\indices{^{a}_{bc}}$, defined through $T\indices{^{a}_{bc}}X^b Y^c = \nabla_b X^b Y^a - X^a \nabla_b Y^b - \left[ X,Y\right]^a$, is a measure of the antisymmetry of a connection: in a coordinate basis, it reads $T\indices{^{\mu}_{\nu\lambda}} = \Gamma\indices{^{\mu}_{\nu\lambda}} - \Gamma\indices{^{\mu}_{\lambda\nu}}$, where $\Gamma\indices{^{\mu}_{\nu\lambda}}$ are the connection coefficients associated to the derivative operator $\nabla$ in this basis. In GR, the connection is metric compatible, in the sense that $\nabla_a g_{bc} =0$ (failure of this condition implies non-metricity, which is the geometric property upon which STGR is built), and torsion-free, in the sense that the associated torsion tensor vanishes. In TPG, by contrast, one uses an alternative so-called `Weitzenb\"{o}ck connection', with torsion but no curvature: see \cite{TPGbook}.

Spacetime curvature constitutes a measure of the extent to which a single vector fails to come back to itself when parallel transported around a loop. Similarly, spacetime torsion constitutes a measure of the extent to which two vectors may fail to form a parallelogram when parallel transported along one another. To see this, take two vectors $\chi^a$ and $\zeta^a$ in the tangent space at some point $p \in M$ where $M$ is the spacetime manifold; first parallel transport $\chi^a$ along  $\zeta^a$, and then transport  $\zeta^a$ along $\chi^a$. In a torsion-free spacetime, the result of these two processes will be the same, and a parallelogram is formed. However, if the connection has torsion, then the `parallelogram' will not close---with this non-closure proportional to torsion.
Given any parallelogram which does not close, one may define therefrom a torsion tensor, and so a connection with torsion.

The Einstein-Hilbert action of GR,
\begin{equation}
    S_{\text{EH}} = \int_M \sqrt{-g} R,
\end{equation}
where $R$ is the Ricci scalar, is equivalent up to a boundary term to the TPG action,
\begin{equation}
    S_{\text{TPG}} = \int_M \sqrt{-g} T,
\end{equation}
where $T$ is the `torsion scalar', which is obtained from the torsion tensor via suitable index contraction.\footnote{See e.g.~\cite{TPGbook} for the explicit definition of the torsion scalar, which won't matter for our purposes.} Since GR and TPG are therefore dynamically equivalent, any empirical phenomenon which one can account for using the resources of one theory can likewise be accounted for using the resources of the other theory. Therefore, insofar as one can account for the full results of a gravitational redshift experiment beyond first order using spacetime curvature in GR, one can likewise account for the full results of such experiments using torsion in TPG. In this sense, curvature is---as already stated above---sufficient but not \emph{necessary} to account for these experimental results.\footnote{For further discussion of the fact that TPG can pass many---in fact, all!---of the `classic tests' of GR, see \cite{WolfRead2}.} This point is not widely known, but deserves to be stressed.

\subsection{Gravitational redshift as evidence for spacetime torsion?}

The conclusion presented above is the correct verdict \emph{vis-\`{a}-vis} other possible geometric explanations of the gravitational redshift results beyond first order. Drawing on work of \cite{schucking2008gravitation}, however, \cite{Maluf} go further, by arguing that gravitational redshift experiments of the Pound-Rebka type provide \emph{direct} evidence for spacetime torsion. This claim cannot be correct; in this subsection, we first present the argument, before diagnosing what is wrong with it.

The argument of \cite{Maluf} proceeds as follows. In a frame comoving with the observers at either end of the Pound-Rebka experimental setup, parallelograms of light rays close---this much is evident from e.g.~Figure \ref{fig:redshift}. But (the reasoning goes) in an inertial frame of reference---\emph{accelerating} with respect to the experimental setup, as already discussed above---such parallelograms do \emph{not} close; therefore, there is direct experimental evidence for spacetime torsion.

This claim is not correct, for several reasons. First, it neglects the above-noted fact that, in an accelerating frame, the two observers will in fact follow the trajectories of \emph{Rindler} observers---recall again Figure \ref{fig:Rindler hyperbolae}. In Rindler spacetime, parallelograms formed by the photons emitted in the experiment \emph{do} close, thereby rendering moot the argument expounded by \cite{Maluf}.

Second---and relatedly---one may always define a collection of vectors which form a `parallelogram' that does not close. However, absent some prior grounding of such a `parallelogram' in the properties of spacetime (e.g.~via the above account regarding the parallel transport of two vectors), to do so is arbitrary, and tells one nothing regarding the nature of spacetime. This, however, is precisely the form of the above argument: a certain `parallelogram' is shown not to close, and an inference regarding spacetime torsion is drawn therefrom. However, no connection exists---or at least, has been shown to exist---between this `parallelogram' and the nature of spacetime: the decision to focus on such a `parallelogram' is arbitrary, with this geometrical construction bearing no relation to e.g.~the parallel transport of vectors about two sides of a loop. Thus (to reiterate), the failure of a parallelogram to close \emph{per se} tells one nothing regarding spacetime torsion.


Third, whether the relevant `parallelogram' closes in the case of gravitational redshift experiments is a manifestly frame-dependent phenomenon. However, whether a spacetime has torsion is a frame-independent matter. The fact that one would construct from this `parallelogram' a vanishing torsion tensor in one frame but not another indicates that one's doing so reveals nothing about the nature of spacetime torsion itself---on the assumption that all facts about spacetime must be frame-independent in nature.



Fourth, at \cite[\S5]{Maluf} it is suggested that the non-closure of the `parallelogram' in gravitational redshift experiments constitutes evidence for TPG (in which the derivative operator has torsion) over GR (in which the derivative operator is torsion-free). However, as already mentioned, the form of TPG under consideration is dynamically equivalent to GR. Thus, it cannot be that \emph{any} empirical results---including those of gravitational redshift experiments---constitute evidence for one theory over the other; and so it cannot be that gravitational redshift results constitute evidence for spacetime torsion. Put another way, even granting that in TPG an explanation for Pounda-Rebka type results can be given in terms of spacetime torsion, it is not the case that such results themselves \emph{favour} TPG torsion-based explanations over alternative, torsion-free explanations available from GR.

We'll close this section with one further observation. In a series of articles, Schild also makes non-trivial inferences about spacetime geometry from considerations to do with parallelograms similar to those of \cite{Maluf} (see \cite[fig.\ 7.1]{Misner:1973prb}); this time, however, the conclusion is that spacetime must be curved! (See \cite{Schild1, Schild2, Schild3}; for a concise summary of Schild's arguments, see \cite[\S7.3]{Misner:1973prb}.) We needn't rehearse here all the details of these arguments---rather, we need only note that the key premise is that ``if flat Minkowski geometry were valid, [...] $\tau_{\text{bot}} = \tau_{\text{top}}$, thus contradicting the observed redshift effect'' \cite[p.\ 189]{Misner:1973prb}. (Here, $\tau_{\text{bot}}$ and $\tau_{\text{top}}$ are two sides of the parallelogram.) Note that Schild insists that it must in fact be the case that spacetime is curved in order for the parallelogram to close, whereas Maluf accepts that the parallelogram does \emph{not} close but introduces spacetime torsion to account for this.
But as we have already seen, neither of these inferences is correct, for there can be a discrepancy between intervals at the top and bottom of a Pound-Rebka setup for accelerating systems even in Minkowski spacetime---i.e.,\ for systems situated in Rindler spacetime!
To repeat, then: to first order, as an explanation of observed gravitational redshift effects, one need invoke neither spacetime curvature nor spacetime torsion.\footnote{We thank an anonymous referee for drawing our attention to Schild's arguments and for pushing us to engage with them.}

\section{Redshift due to charge}\label{sec:charge}

To recap: we've now seen that (a) one needn't invoke geometrical properties of spacetime such as curvature in order to explain first-order gravitational redshift results---here, consideration of accelerating frames in special relativity suffices. Moreover, (b) even beyond first order, one can appeal to other geometric properties of spacetime---\emph{viz}.,~torsion or non-metricity---in order to account for the results of gravitational redshift experiments. In this section, we consider what would be implied by the \emph{absence} of gravitational redshift results: na\"{i}vely, one might think that this would imply that spacetime is Minkowskian; in fact, however, charge in the Reissner-Nordström metric can lead to the cancellation of redshift effects (one might, indeed, be motivated to think this on the grounds that shielding of forces in Reissner-Nordström spacetimes is already a known phenomenon: see e.g.~\cite{Celerier:2017bny}). Therefore, null results of gravitation redshift experiments do not imply that spacetime is Minkowskian.\footnote{There is also the possibility of gravitational redshift in non-relativistic spacetimes---see e.g.~\cite{ObersTests}---but we'll set this aside here.}

\subsection{The weight of photons}

What Pound and Rebka call the `weight of photons' in their experiments, in fact, aptly describes how Einstein originally had thought of gravitational redshift and what he had termed the inertia of energy. 

\subsubsection{A thought experiment}

Let us go back to the thought experiment alluded to in the introduction. Einstein foresaw the gravitational redshift on the basis of a thought experiment using the `inertia of energy' he had discovered in 1905 \cite{einstein1905tragheit}, six years before his famous paper on relativity \cite{einstein1911einfluss}. Here, we'll spell out a variant of this thought experiment (we don't make any historical claim to be reconstructing the argument as Einstein himself presented it).

\begin{figure}[H]
\label{Einstein mass}
\centering
\includegraphics[width=0.9\linewidth]{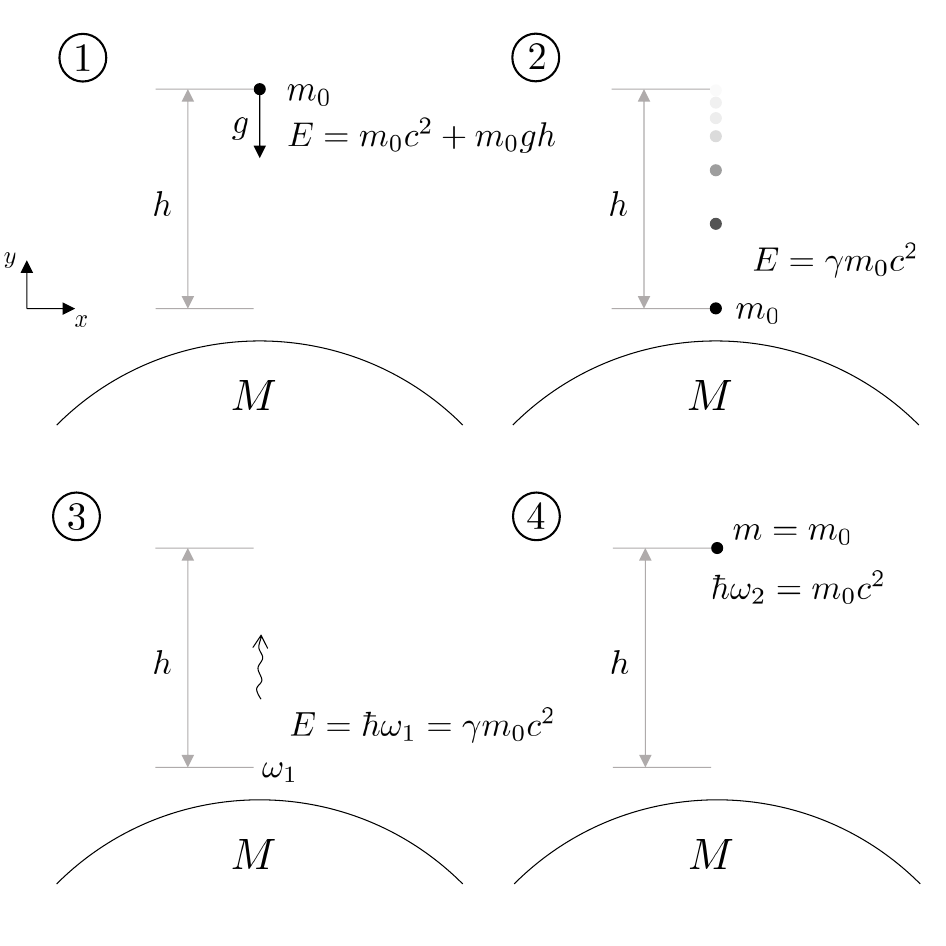}
\caption{Gravitational redshift as a consequence of energy conservation. A test body of mass $m_0$ at rest at a height $h$ is dropped. When it reaches the ground the total energy $\gamma m_0c^2$ is obtained. The mass subsequently is transformed into a photon of energy $\hbar \omega_1$, which is then sent from the ground back to height $h$. There, the photon is transformed back into a mass $m$. By energy conservation, $m$ must equal the mass $m_0$, from which it follows that the photon's frequency must have decreased at its ascent.}
\label{fig:Einstein mass}
\end{figure}

Consider a test body of mass $m_0$ at rest at a height $h$, with a total energy $m_0c^2+m_0gh$---i.e.,~the sum of of its rest energy and gravitational potential energy. Subsequently, the mass is dropped; when it reaches the ground the total energy $\gamma m_0c^2$ is obtained, where $\gamma= \frac{1}{\sqrt{1-\frac{v^2}{c^2}}}$, and $v$ is the velocity of the mass at the ground (such that $m_0c^2+m_0gh=\gamma m_0c^2$). The mass is then transformed into a packet of radiation of energy $\hbar \omega_1$, which is then sent from the ground back to height $h$, where the mass $m_0$ had been situated initially. There, the packet is transformed back into a mass $m$. By energy conservation, $m$ must equal the mass $m_0$ (note that we assume here that the energy of the radiation is transformed entirely into the rest mass of the test body, and not into a sum of rest mass and potential energy), which amounts to saying that $\hbar \omega_2=m_0c^2$, where $\omega_2$ is the frequency of the packet at height $h$. See steps (1)--(4) in Figure \ref{fig:Einstein mass}. 

\noindent From this we regain the first order approximation in \eqref{redshift};
\begin{equation}
\label{redshift_inertia}
\frac{\nu_1}{\nu_2}= \frac{m_0c^2+m_0gh}{m_0c^2}=1+\frac{gh}{c^2}.
\end{equation}  \eqref{redshift_inertia} again involves an approximation of the exact redshift formula, for we assume a uniform gravitational field. Hence we use $m_0gh$ for the energy of the test body. If we were to take into account the $\frac{1}{r}$-dependence of the gravitational potential, then we would obtain
\begin{align}
\frac{\nu_1}{\nu_2}&= \frac{m_0c^2+\int\limits_{r_2}^{r_1}F_Ndr}{m_0c^2}\nonumber\\
&=1+\frac{GM}{c^2}\left(\frac{1}{r_1}-\frac{1}{r_2}\right)\nonumber\\
&\approx1+\frac{gh}{c^2},
\end{align} with $F_N$ being Newton's gravitational force of a massive central body.

Bear in mind that neither the derivation by means of uniformly accelerated frames nor the derivation by means of energy conservation yield the correct value for the gravitational redshift in the first line of \eqref{redshift}. The former holds in virtue of the inhomogeneity of Earth's gravitational field and the merely local validity of the equivalence principle. The latter is true because the Newtonian central body force law is an approximate limit of GR. 

\subsubsection{The inertia of energy}

The approach of describing the redshift effect as a result of energy conservation suggests the following idea:
\begin{quote}
Any `source' of energy causes clocks at different distances from the `source' to exhibit time dilation effects.
\end{quote}
As one example, charged particles attracted by a charged source should likewise be expected to give rise to redshift effects. We can, however, now follow the procedure from above and play the same game with charged bodies, replacing the Newtonian potential with the Coulomb potential. Consider a charged source $Q$ and a test particle of charge $q$ and mass $m_0$. We assume the mass of the source to be negligible. The charged particle falls under the attraction of the source according to the Coulomb force. When it reaches height $r_1$, a photon is created out of it and sent back to the particle's initial position, where it is transformed back into a mass $m$ with charge $q$. For this process to happen, we can imagine annihilating the descending charge by an anti-charge $-q$ to create a photon (or actually at least two photons, which we can think of a single photon for the discussion). The photon is sent back, and when it reaches the top, the initial charge $q$ plus its anti-charge $-q$ is created via pair-production. We assume the two particles have the same mass $m$. The anti-charge $-q$ subsequently is brought back to the bottom to restore the initial situation. It is precisely the energy contribution of this last step that cancels a redshift effect in the calculations, which is not further analysed here.

\subsection{Reissner-Nordström metric}

In fact, charge does give rise to redshift effects---and consequently time dilation---in the standard formalism of GR, albeit not in a way analogous to how mass curves spacetime.

From the Einstein equations, we obtain the Reissner-Nordström metric (cf.~\cite{reissner1916eigengravitation})
\begin{align}
\label{RN}
ds^2=&-\left(1-\frac{2GM}{c^2r}+\frac{GQ^2}{4\pi\varepsilon_0c^4r^2}\right)c^2dt^2\nonumber\\
&+\left(1-\frac{2GM}{c^2r}+\frac{GQ^2}{4\pi\varepsilon_0c^4r^2}\right)^{-1}dr^2\nonumber \\
&+r^2(d\vartheta^2+\sin^2\vartheta d\varphi^2),
\end{align} from which we recover the Schwarzschild metric when $Q=0$. It is worth mentioning that the charge term in the Reissner-Nordström metric affects geodesics of particles even though they may be uncharged.
For $Q\neq 0$, this metric gives rise to an additional gravitational redshift.
In analogy with the derivation of gravitational redshift due to mass, we obtain
\begin{align}
\label{blueshift}
\frac{\nu_1}{\nu_2}&= \frac{\sqrt{\left(1-\frac{2GM}{c^2r_2}+\frac{GQ^2}{4\pi\varepsilon_0 c^4r_2^2}\right)}}{\sqrt{\left(1-\frac{2GM}{c^2r_1}+\frac{GQ^2}{4\pi\varepsilon_0 c^4 r_1^2}\right)}}\nonumber\\
&\approx 1+\frac{gh}{c^2}-\frac{g_{C_2}h}{c^2}, 
\end{align} where $g$ defined as before, and $g_{C_2}:=\frac{GQ^2}{4\pi\varepsilon_0c^2r^3}$. The approximations are as in the case without charge (first order terms in $h$ and large radii $r_1,r_2$). 

The effect is quadratic in the charge $Q$, and, in fact, leads to a \textit{blueshift} of the photon. Thus, it partly compensates the gravitational redshift due to mass. Note that gravity is fully `geometrised' by GR. That is, geodesics of the metric fully describe the motion of test particles. Whereas for charged sources, the usual force terms from elecrodynamics need to be considered additionally in the geodesic equation. 

\subsection{Cancelling redshift with charge}

It is often considered to be a feature of gravity that shielding an object from the influence of a gravitational field is impossible---unlike in e.g.~electromagnetism, where both positive and negative charges exist. But the Reissner-Nordström metric complicates this picture, for at a point in this spacetime one finds that that the charge $Q$ \emph{can} cancel an attractive force towards the black hole.\footnote{Of course, this is subtle, since (a) all particles still move on geodesics, and (b) it really depends on what one means by `gravity'.} This result and its derivation are  already known in the literature as electro-gravitic repulsion---see e.g.~\cite{QADIR1983419}. What is not known is that one can likewise affect gravitational redshift effects due to charge.

Let us elaborate on this point. Recall the Reissner-Nordström metric \eqref{RN}. There, the two terms in $g_{00}$---one proportional to $M$, the other to $Q$---come with opposite signs. This makes it possible to tune the parameters so that Schwarzschildian gravitational redshift effects can be compensated for by the charge of the black hole $Q$, at least locally. Indeed, if we choose the mass and charge such that 
\begin{equation}
\frac{2GM}{c^2r}=\frac{GQ^2}{4\pi\varepsilon_0c^4r^2},
\end{equation} then we recover the Minkowski metric for flat space. This equality, obviously, can only be met on a sphere with one fixed radius $r$. To exactly cancel the redshift effect would require to cancel gravity at two different (and arbitrary points), which is impossible. However, the effects do cancel out to first order, and this might be taken to mean that gravity (at least as manifested in gravitational redshift), in this sense, can at least be `shielded' locally. To be completely clear and to put this point in another way: strictly speaking, the results of this section regarding the cancelling of gravitational redshift with charge are \emph{ultralocal}: they hold exactly only at a single point in spacetime (akin to e.g.\ Lagrange points in the three body problem).\footnote{Strictly speaking, since we're dealing with cancellation of redshift effects by charge at a certain value of the radial coordinate $r$, we in fact have cancellation on an entire spherical shell. However, our results are still ultralocal in the sense that there is no exact cancellation at all points in the neighbourhood of any given point. Other set-ups might in fact be more than merely ultralocal in this sense: consider e.g.\ the cancellation of redshift effects from a uniform gravitational field by an infinite charged plane.} That said, they may hold \emph{approximately} in a broader neighbourhood of that point, insofar as experimental measuring devices cannot detect deviation from absolute shielding (cf.\ `approximate' versions of the strong equivalence principle discussed by \cite{Read2018-REATMO-15}); as such, the considerations presented here remain of operational relevance.

We must be careful to identify correctly the physical significance of the parameter $M$ in the Reissner-Nordström metric. Recall first that typically in the classical limit of a general relativistic spacetime, one writes $g_{00} \cong 1 + 2\phi / c^2$, for an effective Newtonian gravitational potential $\phi$.\footnote{Of course, there are more sophisticated ways in which to take the non-relativistic limit of a model of GR---see e.g.,~\cite{Obers} for recent work in this direction---but here we focus on the `textbook' treatment of that limit.} 

Due to the relativistic equivalence of mass and energy, the electric field energy contributes to the total mass. Taking this into account, the effective total mass $M$ that features in the Reissner-Nordström metric is then found to be

\begin{equation}
    M = M_b +\frac{Q^2}{16\pi \epsilon_0 G M_b},
\end{equation} where $M_b$ is the irreducible bare mass of the black hole (see for instance \cite{CR}. \cite{Damour-RNmass} and \cite{QADIR1983419}).

Thus, the total source mass $M$ in the $g_{00}$ component of the Reissner-Nordström metric, is composed of a term due to the `bare mass' of the black hole, plus a term due to the electric field density. Although the mass term depends on the charge, one can still obtain a cancellation of the redshift effects by charge whenever
\begin{align}
\left(2M_b +\frac{Q^2}{8\pi\epsilon_0G M_b}\right)r=\frac{Q^2}{4 \pi\epsilon_0 c^2}
\end{align}
in which case one sees that one can cancel the gravitational field via the charge $Q$---so, one invariably expects a gravitational \textit{blueshift} effect for small enough radii in the context of the Reissner-Nordström metric. This fits the existence of a repulsive force, since we have already seen that effective forces on test bodies \emph{can} be cancelled using the charge $Q$. The conclusion, then, is that the absence of gravitational redshift effects does not imply a Minkowskian spacetime structure.


\section{Conclusion}\label{sec:conclusion}

In light of this work, what can really be inferred from the results of gravitational redshift experiments? First, if one's experiments (like of those of Pound and Rebka) are insufficiently sensitive, then one is warranted in inferring only the Minkowski spacetime structure of special relativity---for, as \cite{Brown2016} point out, special relativity in accelerating frames is then sufficient to account for these results. Beyond first-order, however, special relativity will not suffice; one might think that in such contexts one must appal to spacetime curvature, but in light of the geometric trinity, this is also incorrect: one could alternatively infer to the existence of spacetime torsion or non-metricity. (\emph{Pace} \cite{Maluf}, however, one cannot infer from these experiments to spacetime torsion uniquely.) Finally, one cannot infer from the absence of gravitational redshift effects to Minkowski spacetime structure, given the possibility of cancelling such effects using charge in Reissner-Nordström spacetimes, and the existence of gravitational blueshift due to charged sources.
Together, we hope that these conclusions will prove definite and final regarding what gravitational redshift experiments really establish.

\section*{Acknowledgements}

Our thanks Harvey Brown, Erik Curiel and Patrick D\"{u}rr for helpful discussions and comments. J.R.~is supported by a Leverhulme Trust Research Fellowship titled `Measuring Spacetime'. 

\bibliography{library}
\end{multicols}
\end{document}